\documentclass[prl,a4paper,twocolumn,showpacs,superscriptaddress]{revtex4}
\usepackage{amsmath}
\usepackage{amsfonts}
\usepackage{amssymb}
\usepackage{bm}
\usepackage{graphicx}
\bmdefine\bomega{\omega}
\bmdefine\bOmega{\Omega}
\bmdefine\bnabla{\nabla}
\bmdefine\bkappa{\kappa}
\bmdefine\bphi{\phi}
\begin{document}

\title{Twisted Vortex State}

\author{V.B.~Eltsov}
\affiliation{Low Temperature Laboratory, Helsinki University of Technology, P.O.Box
2200, 02015 HUT, Finland }
\affiliation{Kapitza Institute for Physical Problems,
Kosygina 2, 119334 Moscow,  Russia}

\author{A.P.~Finne}
\affiliation{Low Temperature Laboratory, Helsinki University of Technology, P.O.Box
2200, 02015 HUT, Finland }

\author{R.~H\"anninen}
\affiliation{Department of Physics, Osaka City University,
Sugimoto 3-3-138, Sumiyoshi-ku, Osaka 558-8585, Japan}

\author{J.~Kopu}
\affiliation{Low Temperature Laboratory, Helsinki University of Technology, P.O.Box
2200, 02015 HUT, Finland }

\author{M.~Krusius}
\affiliation{Low Temperature Laboratory, Helsinki University of Technology, P.O.Box
2200, 02015 HUT, Finland }

\author{M.~Tsubota}
\affiliation{Department of Physics, Osaka City University,
Sugimoto 3-3-138, Sumiyoshi-ku, Osaka 558-8585, Japan}

\author{E.V. Thuneberg} \affiliation{Department of Physical Sciences,
  P.O.Box 3000, 90014 University of Oulu, Finland}

\date{\today}
\begin{abstract}
We study a twisted vortex bundle where quantized vortices form helices
circling around the axis of the bundle in a ``force-free''
configuration.  Such a state is created by injecting vortices into
rotating vortex-free superfluid. Using continuum theory we
determine the structure and the relaxation of the twisted state. This is
confirmed by numerical calculations. We also present experimental
evidence of the twisted vortex state in superfluid
$^3$He-B.
\end{abstract}
\pacs{67.57.Fg, 47.32.-y, 67.40.Vs} \maketitle

The equilibrium state of a superfluid under rotation
consists of an array of quantized vortices, which are parallel to the
rotation axis.  Similarly, the equilibrium state of a type II
superconductor in magnetic field consists of an array of flux
lines parallel to the field. Here we consider {\em twisted vortex
states} where the vortices have helical configuration circling a
common axis. An example is a vortex bundle deformed under
torsion. One type of twisted vortex state appears in a
superconducting current-carrying wire in parallel external magnetic
field \cite{Walmsley,Brandrev,Kamien}. The
current induces a circular magnetic field, which makes the field lines
helical, and in order to be  {\em
force free} the flux lines take the same conformation. Here we
concentrate on a new type of twisted state, which can occur even
when the driving field is a constant. Only this second type can appear
in charge-neutral superfluids, where the rotation is not affected
by currents.

In this letter we demonstrate that the twisted vortex
state appears spontaneously when vortex lines expand
into vortex-free rotating superfluid. We present
analytical results for the twisted state using the continuum theory of
vorticity. In particular, we state the equilibrium
force-free conditions for a uniformly twisted state, and find the
equations governing the relaxation of a
nonuniform twist. We present numerical simulations for
both the generation and the relaxation of the twist.
We discuss the stability of the twisted state, which is limited by
the {\em helical instability} \cite{GJO,Clem}.
The results are valid in superfluids and also in
superconductors in the limit of large penetration depth and no pinning. 
 Finally, we
present experiments that show evidence of the twisted vortex state in
superfluid $^3$He-B.

\begin{figure}[tb]
\centerline{\includegraphics[width=0.5\linewidth]{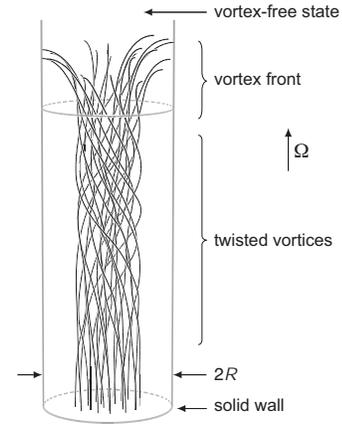}}
\caption{The formation of twisted vortex state. The vortices have
their propagating ends bent to the side wall of the rotating cylinder.
As they expand upwards into the vortex-free state, the ends
of the vortex lines rotate around the cylinder axis. The twist is
nonuniform because boundary conditions allow it to unwind at the
bottom solid wall. The figure gives a snapshot (at time
$t=25\Omega^{-1}$) of a numerical simulation of 23 vortices
initially generated near the bottom end ($t=0$). The parameters are
$2\pi\Omega R^2/\kappa= 86$, $\alpha=0.18$ and $\alpha'= 0.16$
\cite{Bevan}, and
$R/a=3\times10^{5}$, which corresponds to $T$ = 0.4 $T_{\rm c}$ in
$^3$He-B at 29\,bar pressure. For time evolution see Ref.\
\cite{web}. }
\label{f.ekakuva}
\end{figure}

{\it Generation:}---We study superfluid in a long cylinder that
rotates around its axis. We assume that initially the system is in
metastable state, where no vortex lines are present. Then a bunch
of vortex loops is created at some location. They start to expand along
the cylinder. A snapshot from our  numerical simulation of such
propagating vortices is shown in Fig.\ \ref{f.ekakuva}.
Two striking observations can be made. First, the vortices form a front,
where the ends of the lines bend to the side wall. Second, the
growing vortex bundle behind the front is twisted, because the front
rotates at a different speed than vertical vortex sections. The
existence of the front is deduced from simulation and experiment. Here
we concentrate on the twisted state behind the front.

The equilibrium state of the superfluid consists of an array of
rectilinear vortex lines at areal density $n_{\rm v}=2\Omega/\kappa$,
where
$\bm{\Omega}$ is the angular velocity and $\kappa$ the circulation
quantum \cite{hall}. The superfluid velocity 
$\bm{v}_{\rm s}$
at the location of each vortex is
precisely equal to the normal fluid velocity $\bm{v}_{\rm
n}=\bm{\Omega}\times \bm{r}$, so that the array rotates rigidly
together with the cylindrical container. In contrast, the superfluid
velocity vanishes in the vortex-free state, $\bm{v}_{\rm
s}\equiv 0$. All velocities are given in the laboratory frame.

Next we determine the rotation of the vortex ends on the side
wall. For simplicity we consider zero temperature, where the velocity
of a vortex line
$\bm{v}_{\rm L}$ is equal to the local superfluid velocity. The
average superfluid velocity in the vortex front is the average of
$\bm{v}_{\rm s}$ between the vortex state and the vortex-free state.
This gives that the average angular velocity of the front is half
of the container velocity, $\langle\dot\phi\rangle=\Omega/2$. Thus
the vortex front lags behind the vortex array which gives rise to
the twisted vorticity.

It is interesting that the angular velocity of the front can also
be determined from alternative arguments. One is based on the
Hamiltonian equations $\dot\phi=\partial H/\partial L$ and $\dot
L=-\partial H/\partial \phi$, where $L$ is the component of angular
momentum along the cylinder axis $z$. Shifting the vortex front
vertically, the former equation gives $\dot\phi=\Delta E/\Delta
L$, where $\Delta E$ is the energy difference and $\Delta L$ the
angular momentum difference between the vortex and vortex-free
states. Evaluating these using the continuum model of vorticity
($\bm{v}_{\rm s}\approx\bm{\Omega}\times \bm{r}$ in the vortex
state) gives the same result as above. This result is also easy to
generalize to the case where the vortex number $N$ is smaller than
in equilibrium, and the result is
\begin{equation}
\dot\phi=\frac{N\kappa}{2\pi}\frac{\ln(R/R_{\rm
v})+1/4}{R^2-R_{\rm v}^2/2}, \label{e.vorprec}\end{equation} where
$R_{\rm v}^2=N\kappa/2\pi\Omega$. A third argument relies on the
Josephson relation, where the rotating vortex ends cause a phase
slippage to compensate the chemical potential difference between
the two states \cite{volovik}. The rotation of one vortex in
agreement with Eq.\ (\ref{e.vorprec}) has been observed
experimentally by Zieve {\it et al.} in a cylinder with a wire on
the axis and zero applied flow \cite{zieve}.

{\it Uniform twist:}---We construct a description of the twisted
vortex state using the continuum model of vorticity
\cite{hall,BK,sonin}. We start by considering a twisted state which has
translation and circular symmetry. This limits the superfluid
velocity to the form
\begin{equation}
\bm{v}_s=v_\phi(r)\hat{\bm{\phi}}+v_z(r)\hat{\bm{z}},
\label{e.vsutwist}\end{equation} in cylindrical coordinates
$(r,\phi,z)$. It is straightforward to calculate the vorticity
$\bm{\omega} =\bm{\nabla}\times \bm{v}_{\rm s}$. The motion of a
vortex (velocity $\bm{v}_{\rm
L}$) is
determined by the general equation 
\begin{equation} \bm{v}_{\rm L}=\tilde{\bm{v}}_{\rm s} +\alpha
\hat{\bm s} \times (\bm{v}_{\rm n}-\tilde{\bm{v}}_{\rm s})
-\alpha'
\hat{\bm s} \times [\hat{\bm s}
\times (\bm{v}_{\rm n}-\tilde{\bm{v}}_{\rm s})]\,.
\label{e.vl}
\end{equation}
This includes the mutual friction between the vortex lines and the
normal fluid with coefficients $\alpha$ and $\alpha'$. Here
$\hat{\bm s}$ is a unit vector along a vortex line and
$\tilde{\bm{v}}_{\rm s}$ is the superfluid velocity at the vortex core.
In continuum theory
$\hat{\bm s}=\hat{\bm{\omega}}$ (the unit vector along
$\bm{\omega}$), and $\tilde{\bm{v}}_{\rm s}= {\bm{v}}_{\rm
s}+\nu\bm{\nabla}\times\hat{\bm{\omega}}$ differs from the average
velocity $\bm{v}_{\rm s}$ by a term caused by the vortex curvature
\cite{hall}. In our case it is a small correction but is included 
for completeness. Here
$\nu=(\kappa/4\pi)\ln(b/a)$, $b$ is the vortex spacing, $a$ the
core radius.
Note that only the component perpendicular to $\hat{\bm s}$ of
Eq.\ (\ref{e.vl}) is relevant since $\bm{v}_{\rm L}$
parallel to $\hat{\bm s}$ is of no consequence. 

We require that vortices do not move radially in the twisted state
(\ref{e.vsutwist}). This  gives the condition
\begin{equation} (\Omega
r-v_\phi)\left(\frac{v_\phi}{r}+\frac{dv_\phi}{dr}\right)
-v_z\frac{dv_z}{dr}+\frac{\nu}{\vert\omega\vert
r}\left(\frac{dv_z}{dr}\right)^2=0. \label{e.rbr}\end{equation}
Moreover, this condition
implies that all frictional forces vanish since the
twisted vortices rotate uniformly with the cylinder, $\bm{v}_{\rm
L}=\bm{\Omega}\times \bm{r}$.  The only deviation from solid body
rotation on the average is swirling ${\bm{v}}_{\rm s}$ that everywhere
is parallel to the twisted vortices. We conclude that
Eqs.~(\ref{e.vsutwist}) and (\ref{e.rbr}) represent a family of stable
uniformly twisted states. The wave vector
$Q=\omega_\phi/\omega_z r$ of the twist is an arbitrary function of the 
radial coordinate,
$Q(r)$. An explicitly solvable case is obtained by
choosing a constant $Q$ and neglecting $\nu$:
\begin{equation}
v_\phi(r)=\frac{(\Omega+Qv_0)r}{1+Q^2r^2},\ \
v_z(r)=\frac{v_0-Q\Omega r^2}{1+Q^2r^2}.
\label{e.uniksol}
\end{equation}
An important property of the twisted states is the flow parallel to
the axis, $v_z(r)$. Assuming that there is no net flow gives an
integral condition for $v_z(r)$. In the case of Eq.\
(\ref{e.uniksol}) this implies
$v_0=(\Omega/Q)[Q^2R^2/\ln(1+Q^2R^2)-1]$. The deviation of $v_\phi$ from
$\Omega r$ implies that vortices are more compressed in the center
and diluted at larger $r$ compared to equilibrium rectilinear
vortices.

We note that also the Navier-Stokes equations have a stationary
solution for uniform swirling flow, but only under a pressure
gradient along $z$. The twisted state is closely related to the
inertia wave in rotating classical fluids. Various forms of twisted
vorticity as solutions of the Euler equation have been studied in the
literature
\cite{okulov}. 

{\it Nonuniform twist:}---Next we construct equations governing
the relaxation of twisted vortices. Now all components $(v_r,
v_\phi, v_z)$ of $\bm{v}_s$ are nonzero and functions of $r$, $z$,
and time $t$.  Here we take into account only first order
deviations from the rotating equilibrium state. The dynamical
equations can be formed using again Eq.~(\ref{e.vl}) for
$\bm{v}_{\rm L}$, to obtain equations for the radial and azimuthal
coordinates of vortices. These together with the continuity
equation and $\bm{\omega} =\bm{\nabla}\times \bm{v}_{\rm s}$ form
a closed set of equations. The same set of equations has been
derived previously starting from the dynamical equation for
$\bm{v}_{\rm s}$ \cite{GJO,HB}. The essential result is the
dispersion relation \cite{GJO,HB}
\begin{eqnarray}
\frac{(\beta^2+k^2)\sigma}{\Omega}
=-i\alpha(\beta^2+2k^2\eta_2)\nonumber\\
\pm i\sqrt{\alpha^2\beta^4-4(1-\alpha')^2k^2
(\beta^2+k^2)\eta_1\eta_2}\;\;. \label{e.dispr}\end{eqnarray} The
waves giving rise to this dispersion are of the form
$v_r=ckJ_1(\beta r)\exp(ikz-i\sigma t)$ and $v_z=ic\beta J_0(\beta
r)$ $\exp(ikz-i\sigma t)$. $J_0$ and $J_1$ are Bessel functions,
while $\eta_1=1+\nu k^2/2\Omega$ and $\eta_2=1+\nu
(\beta^2+k^2)/2\Omega$.  The boundary condition $v_r(R)=0$ leads
to $\beta= 3.83/R, 7.01/R, \ldots$. We study the case that the
square root in Eq.\ (\ref{e.dispr}) is real.
Then the negative sign corresponds to radial motion of vortices (which
induces also azimuthal motion). Here we are
interested in the positive sign, which corresponds to twisting the
vortex state. In this case the frequency $\sigma$ vanishes for
vanishing wave vector $k$. This is in agreement with our preceding
analysis that there is no relaxation of the uniformly twisted
state. For a nonuniform twist we have to consider a finite $k$,
but still assume $k\ll R^{-1}$. The dispersion relation
(\ref{e.dispr}) then simplifies to $\sigma=-i
k^2(2\Omega/\beta^2+\nu)/d$, where $d$ is another mutual friction
coefficient $d=\alpha/[(1-\alpha')^2+\alpha^2]$. This limit
corresponds to the diffusion equation
\begin{equation}
\frac{\partial f}{\partial t}=D\frac{\partial^2 f}{\partial z^2},\
\ D=\frac{1}{d}\left(\frac{2\Omega}{\beta^2}+\nu\right)
\label{e.diff}\end{equation} with effective diffusion constant
$D$. Here $f(z,t)$ can be either $v_r$ or $v_z$. We see that the
diffusion gets faster towards lower temperatures, where the
friction coefficients approach zero.

\begin{figure}[t]
\centerline{\includegraphics[width=0.99\linewidth]{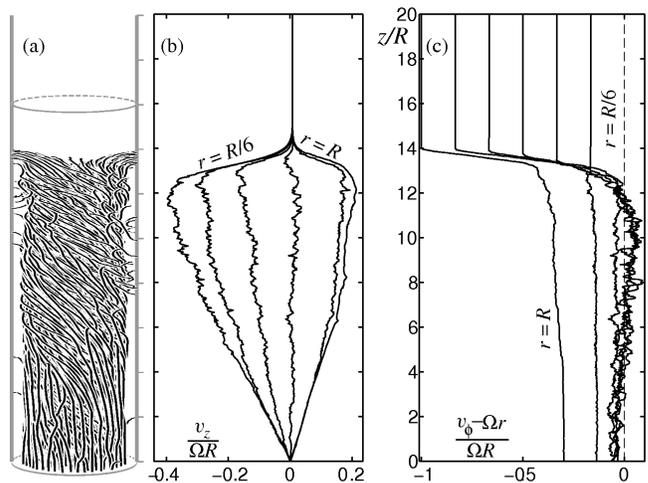}}
\caption{A snapshot of the numerically calculated vortex state expanding
along
$z$: (a) vortex configuration, (b) axial $v_z$, and (c) azimuthal
$v_\phi$ components of the superflow velocity. The velocities, plotted
as functions of $z$, are averaged over the azimuthal angle and are shown
at radii
$r=nR/6$ with integer $n$.  Note that $v_\phi-\Omega r$ changes
sign in (c) close to the center of the bundle, as predicted by
Eq.\ (\ref{e.uniksol}).  The simulation was started
with 203 vortices and the picture was taken after a time interval of
$60\Omega^{-1}$. The parameters are the same as in Fig.\
\ref{f.ekakuva} except $2\pi\Omega R^2/\kappa= 214$,
$R/a=1.5\times10^{5}$.
For clarity,
$r$ and $z$ coordinates have different scales in (a). 
}
\label{f.vz}
\end{figure}

{\it Simulation:}---We have tested the previous theory with
numerical calculations. In the initial
state we have placed a number of vortices at one end of a rotating
cylinder so that they bend from the bottom to the side wall. The
dynamics is  determined by calculating $\tilde{\bm{v}}_{\rm s}$ in
Eq.~(\ref{e.vl}) from the Biot-Savart integral
\cite{sim}. We assume
$\bm{v}_{\rm n}=\bm{\Omega}\times \bm{r}$. An illustrative case
with a small number of vortices is shown in Fig.~\ref{f.ekakuva}.
Another case with more vortices is examined in Fig.~\ref{f.vz}
together with the averaged axial and azimuthal velocity profiles.

The essential features in Fig.~\ref{f.vz} are the vortex free
state in the upper part of the cylinder, the propagating vortex
front, and the twisted vortex state that is left behind. In the
front $v_\phi$ increases rapidly so that the azimuthal counterflow
$v_\phi-\Omega r$ is strongly reduced in absolute value.
While the vortex front progresses, the vortex ends
rotate at a lower speed than the cylinder. This generates the
twisted vortex state. A clear signature of the twist is the axial
velocity $v_z$. It is downwards in the center and upwards in the
periphery (corresponding to a left handed twist, $Q<0$).
At the bottom wall the boundary condition prohibits any axial
flow. This implies that the twist vanishes there. We assume that
the vortex ends can slide with respect to
the bottom wall. Thus the winding generated by the front is
unwound at the bottom.

The calculations indicate that the vortex front deviates from any
equilibrium configuration and probably cannot be described by
simple analytic theory. On the other hand, the relaxing twist
seems to obey  the diffusion equation (\ref{e.diff}). The profile
of $v_z$ in Fig.~\ref{f.vz} can be understood as relaxation
towards a steady state where $v_z$ is linear in $z$.

The twist implies superflow parallel to the vortex lines. In such
a case it is expected that individual vortices can become unstable
against helical distortion. A calculation in Ref.~\cite{GJO}
predicts this {\em helical instability} to take place for rectilinear
vortices when the velocity of the parallel flow reaches
$v_z=2\sqrt{2\Omega\nu}$. The simulations indicate that the
maximum axial velocity $v_z$ (see Fig.~\ref{f.vz}) remains smaller
than this limit. It appears that if any tighter twist is created in the
front, it is immediately relaxed by instabilities and subsequent
vortex reconnections. Note that the same helical instability is
responsible for flux flow in the force-free configurations
in superconductors \cite{Clem,Brandrev}. 

\begin{figure}[tb]
\centerline{\includegraphics[width=0.9\linewidth]{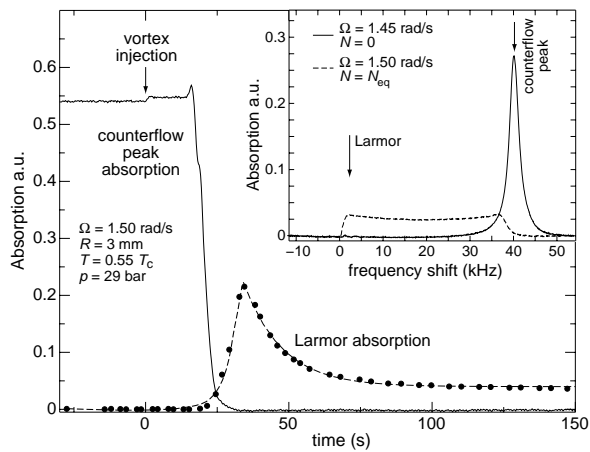}}
\caption{NMR absorption signals as a function of time after
injection of vortices at $t=0$. The two NMR
absorption records in the {\it main panel} show the reduction in the
counterflow peak and the overshoot in the Larmor region.  The former is
interpreted as the arrival of the vortex front and the rapid increase in
$v_\phi$. The latter arises from the axial flow velocity $v_z$, caused
by twisted vortices, and the subsequent slow relaxation towards the
equilibrium state. The {\it insert} shows the NMR line shapes in the
initial vortex-free state $N = 0$ and the final equilibrium vortex
state $N = N_{\rm eq}$. The spectra are measured at constant
temperature and have the same integrated absorption.} \label{f.exp}
\end{figure}

{\it Experiment:}---We now turn to the evidence for the twisted
vortex state from NMR measurements on a rotating sample of
$^3$He-B. The experimental details have been described in
Ref.~\cite{Exp}. What is essential is that the NMR absorption is
measured at two symmetric locations near the ends of the long
sample cylinder. A measuring run is shown in Fig.\ \ref{f.exp}.
The vortices are injected in the middle of the cylinder at time
$t=0$. Then the NMR line shapes change in both locations
simultaneously from the initial vortex-free form ($N=0$ spectrum
in the inset of Fig.~\ref{f.exp}) to that of the final equilibrium
vortex state ($N=N_{\rm eq}$ spectrum). During the transition the
absorption is first shifted from a ``counterflow'' peak to an
overshoot in the ``Larmor region'' and later redistributed more
evenly over the entire spectrum. By tuning one spectrometer on the
counterflow peak and the other on the peak in the Larmor region,
the timing of the two peaks is resolved in the main panel. We see
that after a flight time of 22\,s, the vortex fronts reach the
spectrometers. The spectrometer tuned to the counterflow peak sees
a rapid drop in absorption. The other spectrometer tuned to the
Larmor peak records first a rapid increase, followed by a slow
exponential relaxation with a time constant of 14\,s towards the
final level of the equilibrium vortex state.

The quantitative interpretation of the two signals requires
detailed analysis of order parameter textures in $^3$He-B with
curved vortices and will be presented elsewhere. Very simply,
though, a large counterflow peak comes from counterflow
$\bm{v}_{\rm n}-\bm{v}_{\rm s}$ that is perpendicular to the
rotation axis $z$ (when the static magnetic field is along $z$).
Conversely, large absorption near the Larmor frequency comes from
counterflow parallel to $z$. Because $z$ is a symmetry direction,
there is always some absorption near the Larmor frequency, but it
is modest under normal circumstances, where no axial flow is
present ($N=N_{\rm eq}$ spectrum in Fig.~\ref{f.exp}).

The transient absorption traces in Fig.~\ref{f.exp} can now be
understood as a measurement of the vortex state in Fig.~\ref{f.vz}
at a fixed detector location $z_{\rm det}$. The arrival of the
vortex front at $z_{\rm det}$ causes an abrupt reduction in the
counterflow peak as $\vert v_\phi-\Omega r\vert$ is reduced.
Simultaneously $v_z$ is increasing which is seen as a rise in the
Larmor absorption. The subsequent decrease of $v_z$ after the
passage of the front is seen as relaxation of the Larmor
absorption towards the equilibrium vortex state.

\begin{figure}[tb]
\centerline{\includegraphics[width=0.7\linewidth]{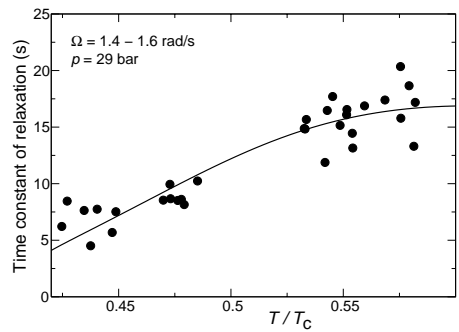}}
\caption{The time constant for the decay of the Larmor absorption (data
  points) compared to the decay of the twisted state according to the
  slowest mode of Eqs.~(\ref{e.dispr}) or (\ref{e.diff}) (line). The line
  has $k = \pi / \ell$ as fitting parameter giving $\ell = 18$ mm.  For
  comparison, the half-length of the cylinder is 55 mm and the detector
  coil is placed between 5 and 15 mm distance from the end plate
  \cite{Exp}. } \label{f.exp2}
\end{figure}

The time constant for the decay of the Larmor absorption is
plotted in Fig.~\ref{f.exp2} as a function of temperature together
with the slowest mode from the diffusion equation (\ref{e.diff}).
We note that the theoretical eigenvalue is in order of magnitude
agreement with the measurements. It is especially noteworthy that  
relaxation gets faster with decreasing temperature. This may at first
seem surprising since the relaxation is usually associated with
friction, which decreases with decreasing temperature. Our simulations,
which are time consuming at the experimental parameter values, yield
a time constant which is larger but within a factor of 3 from the
experimental value. 

In conclusion, in rotating superfluid a front followed by a twisted vortex
bundle is the preferred configuration of vortex expansion, rather than a
turbulent tangle, when the induced supercurrents remain below the helical
instability limit. It appears feasible that a similar state can be
generated in a superconductor in the superclean limit.

We thank N. Kopnin, E. Sonin, and G. Volovik for useful
comments.



\vspace{-5mm}
\begin{thebibliography}{99}

\vspace{-3mm}

\bibitem{Walmsley} D.G. Walmsley, J. Phys. F {\bf 2}, 510 (1972) 

\bibitem{Brandrev} E.H. Brandt, Rep. Prog. Phys. {\bf 58}, 1465 (1995).

\bibitem{Kamien} R.D. Kamien, Phys. Rev. B {\bf 58}, 8218 (1998).

\bibitem{GJO} W.I. Glaberson, W.W. Johnson, and R.M.
Ostermeier, Phys. Rev. Lett. {\bf 33}, 1197 (1974).

\bibitem{Clem} J.R. Clem, Phys. Rev. Lett. {\bf 38}, 1425 (1977).

\bibitem{Bevan} T.D.C. Bevan {\it et al.}, J. Low Temp. Phys. {\bf 109}, 423 (1997).

\bibitem{web} See site: http://ltl.tkk.fi/research/theory/twist.html

\bibitem{hall} H.E. Hall, Adv. Phys. {\bf 9}, 89 (1960).

\bibitem{volovik} T.Sh. Misirpashaev and G.E. Volovik,
Pis'ma Zh. Eksp. Teor. Fiz. {\bf 56}, 40 (1992)
[JETP Lett. {\bf 56}, 41 (1992)].

\bibitem{zieve} R.J. Zieve {\it et al.},
Phys. Rev. Lett. {\bf 68}, 1327 (1992).

\bibitem{BK} I.L. Bekarevich and I.M. Khalatnikov, J. Eksp. Theor.
Phys. {\bf 40}, 920 (1961) [Sov. Phys. JETP {\bf 13},643 (1961)].

\bibitem{sonin} E.B. Sonin, Rev. Mod. Phys. {\bf 59}, 87 (1987).

\bibitem{okulov} G.K. Batchelor,
{\it An introduction to fluid dynamics} (Cambridge, 1988); V.L.
Okulov, J. Fluid Mech. {\bf 521}, 319 (2004).

\bibitem{HB} K.L. Henderson and C.F. Barenghi, Europhys. Lett. {\bf
67}, 56 (2004).

\bibitem{sim} R. H\"anninen{\it et al.}, J. Low Temp. Phys. {\bf 138},
589 (2005).

\bibitem{Exp} A.P. Finne {\it et al.}, J. Low Temp. Phys. {\bf 136}, 249 (2004).


\end{thebibliography}
\end{document}